# Note de synthèse sur l'indice de réfraction de l'air en fonction de la température, la pression, l'humidité et l'ionisation

*Short review on the refractive index of air as a function of temperature, pressure, humidity and ionization*

**Luc Dettwiller** [a]

[a] *Université Jean Monnet Saint-Etienne, CNRS, Institut d Optique Graduate School, Laboratoire Hubert Curien UMR 5516, F-42023, SAINT-ETIENNE, France*
E-mail: dettwiller.luc@gmail.com

**Abstract.** The empirical law of Gladstone-Dale is insufficient for high-precision studies using the refractivity of a gas: this is not exactly proportional to its density, and the gas may not be properly described as perfect. An optical Mariotte temperature allows making a comparative analysis of the results given by various authors. The effect of hygrometry on the refractivity at visible wavelengths is historically traced and its small effect on the astronomical refraction angle numerically shown. Finally at infrared and radio wavelengths, the effects of the humidity in the lower atmosphere can be strong; as for the ionosphere, its curvature plays an essential role for the astronomical refraction angle unlike in the visible.

**Key words.** Air refractive index, air refractivity, Gladstone-Dale rule, Lorentz-Lorenz law, Mariotte temperature, optical Mariotte temperature, water vapour, astronomical refraction, ionospheric refraction.

**Résumé.** On rappelle que la loi empirique de Gladstone-Dale est insuffisante pour des études de précision utilisant la réfractivité d'un gaz : celle-ci n'est pas exactement proportionnelle à sa masse volumique, et cette complication se combine avec les écarts à la loi des gaz parfaits, ce qui nous pousse à introduire une température de Mariotte optique, que nous utilisons dans notre analyse comparative des résultats de divers auteurs. On retrace l'historique de l'étude de l'effet de l'hygrométrie sur la réfractivité dans le visible, et on regarde numériquement ses faibles conséquences sur l'angle de réfraction astronomique. On envisage cet effet dans l'infrarouge, et surtout dans les ondes radio : pour elles les variations de réfraction dans la basse atmosphère sont très fortes ; quant à l'ionosphère, elle produit un angle de réfraction non nul, mais en raison de sa courbure essentiellement – au contraire de ce qui se produit avec l'atmosphère dans le visible.

**1. Introduction**

La connaissance précise de l'indice de réfraction $n$ de l'air (ou de sa réfractivité $\eta = n-1$) est une question de grande importance : d'abord, historiquement, pour la connaissance de la réfraction astronomique pour laquelle nous rédigeons cette étude (voir le numéro spécial des C. R. Phys. intitulé *Astronomie, atmosphères et réfraction*) ; ensuite, pour la spectroscopie de précision (où il faut convertir les longueurs d'onde $\lambda$ mesurées dans l'air en longueurs d'onde dans le vide $\lambda_0 = n\lambda$) ; pour la télémétrie laser – avec la mesure précise de la distance Terre-Lune, etc. L'influence de $\lambda_0$ est bien documentée (voir par exemple [1] § 6.2) ; mais il y a aussi d'autres facteurs, qui n'agissent pas de la même façon selon les domaines spectraux.

Le but du présent article est de fournir un compendium sur ce sujet, en distinguant les phénomènes qui s'y rapportent. Il y a d'abord les variations de $\eta$ avec la température absolue $T$ et la pression $P$ ; nous montrons qu'il est intéressant de synthétiser celles-ci en introduisant une température de Mariotte optique $t'_M$ en degrés Celsius (différente de la température de Mariotte $t_M$ bien connue en degrés Celsius dans l'étude des propriétés thermodynamiques des gaz), et nous comparons les effets sur $\eta$ des corrections aux lois des gaz parfaits et de Gladstone-Dale. Ensuite, viennent les variations de $\eta$ avec l'hygrométrie ; nous retraçons l'historique de leur étude, et nous en donnons des valeurs numériques que nous appliquons à l'étude de la réfraction astronomique [1]. Enfin, nous soulignons l'effet de l'humidité de l'air dans l'infrarouge, et surtout dans le domaine des ondes radio, où il n'est plus possible de relier $\eta$ uniquement à la masse volumique $\mu$ de l'air, contrairement à ce qui se fait avec une très bonne exactitude aux petites longueurs d'onde, jusqu'à l'infrarouge. Nous déduisons de ce fait quelques conséquences axées sur la réfraction astronomique, qui reste une des motivations actuelles de cette étude : d'une part, nous expliquons les deux raisons majeures de sa forte variabilité en ondes radio dans la basse atmosphère ; d'autre part, nous rappelons les lois générales de la réfraction de ces ondes dans l'ionosphère.

Quelques définitions conventionnelles (celles de l'ancien air normal, du nouvel air normal, de l'air standard) nous aideront à entrer dans le vif du sujet :
- l'ancien air normal est par définition de l'air sec, à 15 °C, 1 atm, avec 330 ppmv de $CO_2$ [2,3] ;
- le nouvel air normal est sec aussi, mais à 20 °C, 1 bar, et avec 400 ppmv de $CO_2$ [4] ;
- Ciddor [5-7] adopte l'« air standard » défini par Birch et Downs [8], comme étant l'ancien air normal, mais avec 450 ppmv de $CO_2$.

## 2. Lois précises de variation de la réfractivité avec la température et la pression

Pour des études contemporaines et précises de la réfractivité de l'air, on commence par spécifier minutieusement sa composition ; des formules empiriques permettent de tenir compte des variations de l'hygrométrie et de la fraction molaire de $CO_2$ (les valeurs numériques que nous donnons ci-dessous sont pour l'air sec avec 400 ppmv de $CO_2$). Ensuite, on ne part pas de la loi de Gladstone-Dale reliant linéairement la masse volumique $\mu$ à $\eta$ (pour une composition donnée), mais de la relation de Lorentz-Lorenz (voir [9] chap. IV, [10]), selon laquelle $\mu$ est proportionnel à $\frac{n^2-1}{n^2+2}$ (la constante de proportionnalité ne variant qu'avec $\lambda_0$, mais pas avec l'état physique).

Il est facile d'inverser cette relation [11], et d'en déduire une expression de $(n-1)$ en fonction de $\mu$, non-linéaire (quoique les deux s'annulent en même temps) ; c'est la seconde méthode de Ciddor [5] pour exprimer l'indice de l'air humide, en fonction de son degré hygrométrique. Mais pour l'air sec, au lieu de cela on trouve plus souvent dans la littérature [4,9,12-14] une approximation basée sur une approche itérative. Elle commence par un développement du premier membre de la relation de Lorentz-Lorenz à l'ordre 2 en $\eta$, donc on part de

$$\frac{n^2-1}{n^2+2} \cong \frac{2(n-1)}{3}\left(1-\frac{n-1}{6}\right) ; \qquad (1)$$

on corrige alors la loi de Gladstone-Dale – qui assimile, pour une composition donnée, $\eta(T,P,\lambda_0)$ à $C(\lambda_0)\,\mu(T,P)$ – en écrivant la réfractivité sous la forme

$$\eta(T,P,\lambda_0) \cong \left[1-\frac{\eta(T,P,\lambda_0)}{6}\right]^{-1} C(\lambda_0)\,\mu(T,P), \qquad (2)$$

et pour expliciter le terme entre crochets de la relation (2), on se contente de l'approximation (3), i.e. celle de Gladstone-Dale et des gaz parfaits, en négligeant en plus la dispersion de l'air sur le visible car la variation relative de $C \cong 2{,}3.10^{-4}\,\text{m}^3.\text{kg}^{-1}$ n'y est que de l'ordre de $10^{-2}$ (et cela laissera séparées les variables $\lambda_0$ et $(T,P)$ dans l'expression de $\eta$) :

$$\eta(T,P,\lambda_0) \cong C\mu \cong C\frac{M_a}{R}\frac{P}{T_0}\frac{T_0}{T} = \frac{M_a C}{RT_0}\frac{P}{1+(t/T_0)} \qquad (3)$$

où $t := T - T_0 := T - 273{,}15$ K est la température Celsius (le symbole := indiquant une notation ou une définition).

L'avantage de la séparation des variables $\lambda_0$ et $(T,P)$ pour l'air sec est que son pouvoir dispersif, i.e. $\frac{n_F - n_C}{n_d - 1}$ où $n_F$ et $n_C$ désignent les indices respectifs pour les raies F bleue (à $\lambda_0 \cong 486$ nm) et C rouge de l'hydrogène (à 656 nm), et $n_d$ celui pour la raie d (de l'hélium, à 588 nm), est indépendant de $(T,P)$ ; ceci a été expérimentalement corroboré avec précision en 1960 par Svensson pour la variable $T$, et en 1961 par Erickson [15] pour $P$. Ce n'est plus vrai quand on traite du cas de l'air humide [3-6,12,14-16], car les courbes spectrales de réfractivité de l'air sec et de la vapeur d'eau ont des formes différentes, et on prend comme quatrième variable le degré hygrométrique ou la pression partielle de vapeur d'eau ; mais quand cette quatrième variable est fixée, le titre molaire $x_v$ de vapeur d'eau dans l'air varie avec $(T, P)$. Le problème est souvent simplifié avec $CO_2$ : comme on utilise son titre molaire $x_c$ dans l'air [3-5,16,17], et qu'on néglige (sauf Owens [14], et Ciddor [7]) la différence de forme entre les courbes spectrales de réfractivité de l'air sec (sans $CO_2$) et du $CO_2$ gazeux pur [18], on pourra considérer que dans l'expression de $\eta$ les variables $\lambda_0$ et $(T, P, x_c)$ sont séparées [3-5,16,17].

Maintenant, pour exprimer $\mu(T, P)$ dans (2), on ne se contente plus de l'approximation des gaz parfaits ; on utilise ce qui sert notamment à préciser l'allure des isothermes de l'air en coordonnées d'Amagat $(P, P/\mu)$ au voisinage de l'axe vertical des $P/\mu$ [19] : le développement du viriel, c'est-à-dire celui de la grandeur sans dimension $z := M_a \dfrac{P/\mu}{RT}$, où $M_a \cong 28{,}968$ g.mol$^{-1}$ est la masse molaire de l'air, et $R \cong 8{,}3145$ J.K$^{-1}$.mol$^{-1}$ (qui a une valeur exacte depuis le 20 mars 2019, suite à la révision du système international d'unités) est la constante des gaz parfaits ; $z$ est appelé « (facteur de) compressibilité » (à ne pas confondre avec les compressibilités isentropique ou isotherme, qui sont homogènes à l'inverse d'une pression). En utilisant la température Celsius $t$, Bönsch et Potulski mettent [4], comme développement de $z$ à l'ordre 1 en $P$, l'écriture simplifiée

$$z(T, P) \cong 1 - (a - bt) P \tag{4}$$

car le coefficient de $P$ est approximé par son développement d'ordre 1 en $t$. Ce coefficient s'annule par définition à la température de Mariotte $t_M$ (ou température métacritique) [19], qui vaut $56{,}40$ °C pour l'air sec d'après les valeurs de $a$ et $b$ données ci-dessous.

Ainsi, en simplifiant (3) par un développement d'ordre 1 en $t/T_0$, et en utilisant (2) et (4), on en déduit une expression de la réfractivité plus précise que (3) :

$$\begin{aligned}\eta(T, P, \lambda_0) &\cong \left[1 + \frac{\eta(T, P, \lambda_0)}{6}\right] \frac{M_a C(\lambda_0)}{RT_0} \frac{P}{1 + (t/T_0)} \frac{1}{z(T, P)} \\ &\cong \left\{1 + \frac{M_a C}{6 RT_0}\left[1 - (t/T_0)\right] P + (a - bt) P\right\} \frac{M_a C(\lambda_0, M)}{RT_0} \frac{P}{1 + (t/T_0)}\end{aligned} \tag{5}$$

où le terme entre accolades synthétise les corrections approximatives par rapport aux lois de Gladstone-Dale et des gaz parfaits ; numériquement [4]

$$\eta(T, P, \lambda_0) \cong \left[1 + \frac{0{,}5953 - (0{,}009\,876\, t/\text{°C})}{10^8 \text{ Pa}} P\right] \frac{M_a C(\lambda_0, M)}{RT_0} \frac{P}{1 + (0{,}003\,6610\, t/\text{°C})} \tag{6}$$

car $a \cong 0{,}547\,05.10^{-8}$ Pa$^{-1}$ et $b \cong 0{,}009\,699.10^{-8}$ Pa$^{-1}$.°C$^{-1}$. La seconde correction (celle donnée par $z$) est d'ailleurs plus importante que la première (qui n'a été détectée expérimentalement que plus tard) : pour le nouvel air normal, la correction relative de la réfractivité – par rapport au calcul fait avec l'hypothèse du gaz parfait – vaut $(1/z) - 1 \cong (a - bt) P \cong +0{,}035$ % indépendamment de $\lambda_0$ ; la correction relative de la loi de Gladstone-Dale la renforce un peu – car dans les mêmes conditions, et en plus pour la raie d, elle vaut $\dfrac{\eta(T, P, \lambda_0)}{6} \cong +0{,}0049$ %, ce qui est un ordre de grandeur au-delà de la précision requise selon Stone [20]. On voit que suivant les formules contemporaines précises [4,5,12-14] la réfractivité de l'air sec et sans $CO_2$ n'est pas proportionnelle à $P/T$, ni même à $\mu$, mais à un « facteur de densité » (fonction de $T$ et $P$ seulement), dont le produit par le « facteur de dispersion » (fonction de $\lambda_0$ seulement) donne $\eta$ ; mais on comprend aussi que la séparation des variables $\lambda_0$ et $(T, P)$ dans l'expression de $\eta$ n'est cependant qu'une approximation qu'Edlén adopte consciemment [2].

Il est alors intéressant de définir une température de Mariotte optique $t'_M$, où le coefficient de $P$ dans le développement en $P$ de $\eta(T,P,\lambda_0)T/P$ s'annule. A priori $t'_M$ dépend de $\lambda_0$ (contrairement à $t_M$) ; si l'examen du terme entre crochets dans la relation (6) ne le montre pas, c'est parce que la variation de $C$ avec $\lambda_0$ a été négligée dans l'expression du terme entre crochets de la relation (2). Numériquement, $t'_M \cong 60,28$ °C d'après (6) ; mais si on conserve sans le développer le terme $1/[1+(t/T_0)]$ de (3), on obtient une expression plus compliquée. En utilisant les valeurs numériques fournies par Bönsch et Potulski [4] pour l'air sec avec 400 ppmv de $CO_2$ on trouve alors

$$\eta(T,P,\lambda_0)\frac{T}{P} \cong \left[1 + \frac{16\,256\text{ K} - 210,22\,t - 0,9699\text{ K}^{-1}\,t^2}{10^{10}\text{ Pa}\,(273,15\text{ K}+t)}P\right]\frac{M_a\,C(\lambda_0)}{R}, \quad (7)$$

ce qui donne $t'_M \cong 60,46$ °C (0,3 % au-dessus de la valeur approximative précédente) ; et en utilisant les valeurs numériques de Ciddor [5],

$$\eta(T,P,\lambda_0)\frac{T}{P} \cong \left[1 + \frac{15\,812,3\text{ K} - 293,31\,t + 1,1043\text{ K}^{-1}\,t^2}{10^{10}\text{ Pa}\,(273,15\text{ K}+t)}P\right]\frac{M_a\,C(\lambda_0)}{R} \quad (8)$$

d'où $t'_M \cong 75,202$ °C. La différence de signe entre les coefficients de $t^2$ dans les relations (7) et (8) vient de la simplification due à Bönsch et Potulski et mentionnée ci-dessus concernant le coefficient de $P$ dans le développement de $z$ à l'ordre 1 en $P$, alors que les expressions de Ciddor viennent de la même source [21] mais sans simplification pour ce coefficient.

### 3. Influence de la vapeur d'eau sur l'indice de l'air

En 1816-1818, Fresnel et Arago ont utilisé un interféromètre à deux voies portant chacune un tube long de 1,008 m, l'un rempli d'air humide à saturation, l'autre d'air qu'ils faisaient dessécher progressivement grâce à de la potasse ou du chlorure de calcium dans le tube ; ils ont constaté que cela engendre un déplacement des franges d'interférences de 1,25 interfrange pour $\lambda_0 \cong 577$ nm ([22] p. 719) – d'où une variation d'indice de l'ordre de $0,72.10^{-6}$ – et que « Les franges marchent du côté de l'air sec. L'air humide est donc le moins réfringent. » ([23] p. 332) Plus tard Arago utilisera aussi un « verre rouge » donnant $\lambda_0 \cong 623$ nm ([22] p. 722). La mesure du déplacement des franges était affinée par l'utilisation de lames inclinables jouant le rôle d'un compensateur.

En 1852, Arago a voulu rendre cette expérience plus précise, grâce à un interféromètre ayant des bras de 10 m de long qu'il fait installer dans la salle méridienne de l'Observatoire de Paris : mais, devenu quasiment aveugle à cause du diabète, il a confié les mesures à Fizeau ([22] p. 724).

En 1858 Jamin reprend cette expérience avec des tubes de 4 m [24].

Le dépouillement de ces expériences montre que la réfractivité de la vapeur d'eau dans le visible est (contrairement à une idée préconçue) seulement 0,9 fois celle de l'air sec sous les mêmes conditions de $T$ et $P$ – alors qu'elle est 1,4 fois celle de l'air sec pour une même masse volumique, l'écart venant du rapport des masses molaires. À température $T$, pression *totale* $P$ et titre molaire $x_c$ de CO$_2$ *fixés*, une variation $\Delta P_v$ de pression *partielle* de vapeur d'eau provoque une petite variation de signe opposé pour l'indice de l'air, quasi indépendante de $(T, P, x_c)$ [4] :

$$\Delta n \cong -(\Delta P_v/1\ \text{Pa})(3{,}8020 - 0{,}0384\ \mu\text{m}^2\ \lambda_0^{-2}).10^{-10}.\tag{9}$$

Pour atteindre sur la valeur de la réfraction astronomique la précision de 0,04 % requise en astrométrie fine, il faut donc mesurer la température $T$, la pression totale $P$ et l'humidité relative à 0,1 °C, 0,5 mbar et 0,1 près !

À 20 °C et 1 atm, quand le degré hygrométrique $h$ varie de 0 à 100 %, la pression partielle de vapeur d'eau croît de 23,4 hPa (qui est la pression de vapeur saturante de l'eau à 20 °C), et (à 588 nm) la réfractivité passe alors de 2,724 . 10$^{-4}$ à 2,716 . 10$^{-4}$, soit une variation de – 0,3 % ; la variation de réfraction horizontale est alors – 7″ environ, et – 0,17″ à 45° de hauteur. On voit bien que les variations de réfraction relevées par Delambre ([25] p. 326) et mentionnées au paragraphe 5 de [1] ne sont pas dues à la vapeur d'eau, qui peut être négligée pour les applications usuelles ou anciennes [23,26] – mais pas pour l'astrométrie fine actuelle.

### 4. Cas de l'infrarouge et des ondes radio

L'influence de la vapeur d'eau discutée au paragraphe précédent se rapporte au domaine visible. Les fortes bandes d'absorption de l'eau dans l'infrarouge y provoquent, en vertu des relations de Kramers-Kronig, une forte dispersion, comme le montrent les références suivantes [27-29]. Ces travaux pointent aussi le fait que la valeur plus basse de la réfractivité de l'air dans la fenêtre de transparence de l'air humide (vers 10 μm) est due à la forte absorption à partir du niveau fondamental de CO$_2$ vers 15 μm (conformément au profil de réfractivité classique de part et d'autre d'une raie), et à la bande de rotation pure de la vapeur d'eau à plus grande longueur d'onde.

Dans les ondes radio, la vapeur d'eau a une réfractivité bien plus grande que dans le visible à $T$ et $P$ fixés, donc l'influence de la vapeur d'eau est beaucoup plus forte. Il est bien connu des opérateurs radar que dans leur domaine spectral, le coefficient de réfraction de l'atmosphère (voir le sous-paragraphe 7.1 de [30]) vaut $1/4$ en moyenne, au lieu d'environ $1/6$ dans le domaine visible. Pour les ondes radar ce coefficient est aussi plus variable avec la météo, d'où de fréquentes anomalies. Ceci vient en partie de l'importance de la polarisabilité d'orientation de l'eau, bien augmentée dans le domaine micro-onde (alors qu'elle est insensible dans le visible) à cause de sa raie de rotation pure R(0) qui est celle de plus grande longueur d'onde (1,35 cm) et qui joue donc un rôle important dans l'absorption des micro-ondes par l'atmosphère ; corrélativement, la différence de population des états correspondants à R(0) varie facilement avec la température au voisinage de l'ambiante, ce qui rend la réfractivité de l'air *humide* plus sensible à la température en micro-ondes que dans le visible – mais cet effet est secondaire. Essen et Froome donnent [31] la formule suivante, valable à 24 GHz :

$$10^6 \eta \cong \frac{103{,}49\ \text{K}}{T}\frac{P_1}{1\ \text{torr}} + \frac{177{,}4\ \text{K}}{T}\frac{P_2}{1\ \text{torr}} + \frac{86{,}26\ \text{K}}{T}\left(1 + \frac{5\,748\ \text{K}}{T}\right)\frac{P_3}{1\ \text{torr}}\tag{10}$$

où $P_1$, $P_2$ et $P_3$ désignent respectivement les pressions partielles de l'air pur, du dioxyde de carbone et de la vapeur d'eau. La polarisation d'orientation de l'eau rend, pour sa vapeur, la réfraction molaire et $C$ (la constante de la loi de Gladstone-Dale) nettement dépendantes de la température, comme le montre le terme $\dfrac{5\,748\ \text{K}}{T}$ : il n'est plus possible de relier $n$ uniquement à la masse volumique $\mu$ de l'air.

Une autre cause de la forte valeur et variabilité du coefficient de réfraction en ondes décimétriques est la faible échelle de hauteur $H_v$ de la vapeur d'eau atmosphérique (elle ne suit pas la loi de Boltzmann, et ce n'est pas un gaz bien mélangé à l'air) : $H_v$ est de l'ordre du huitième de la hauteur d'échelle de l'atmosphère, donc l'augmentation de la polarisabilité de la vapeur d'eau accroît le gradient de $n$ huit fois plus que le ferait la même augmentation pour l'air sec (à concentration identique). Cet effet se trouve amplifié lors des très fortes inversions de température au-dessus de l'océan, quand une couche d'air froid et humide au-dessus de l'eau est surmontée d'une couche d'air chaud et sec (comme celui du désert apporté par les vents Santa Ana, ainsi nommés d'après une montagne et un canyon californiens, pouvant produire un gradient thermique vertical de $300\,\mathrm{K\cdot km^{-1}}$ sur 50 m d'épaisseur). Les effets de courbure des rayons en ondes centimétriques sont alors très forts.

Bien sûr, les corrections de réfraction sont prises en compte en radioastronomie et finement étudiées ; un symposium international y a été consacré en 1979 [32]. En ce qui concerne l'influence de l'ionosphère, le tracé des rayons pour les ondes radio a été très étudié, d'abord dans le modèle d'une couche ionosphérique à symétrie sphérique. À cause de l'importance de l'effet des électrons libres (de population $N_e$) sur l'indice *de réfraction* à des fréquences juste supérieures à celle de coupure ionosphérique (de l'ordre de 20 MHz), $\eta < 0$ varie en sens inverse de $N_e$ (qui caractérise l'ionisation), car dans ce domaine $\eta$ est quasi proportionnel à $-N_e\lambda_0^2$ ; dans les années 1950 on savait déjà qu'en conséquence, dans une couche ionosphérique et à ces fréquences, les rayons radio obliques ont leur concavité tournée vers le haut (resp. bas) en dessus (resp. en dessous) de la surface de $N_e$ maximal, contrairement à ce qui se passe pour les rayons lumineux obliques du domaine optique, qui ont toujours leur concavité tournée vers le bas sur toute l'étendue de l'atmosphère normale. Des rayons critiques peuvent présenter un cercle asymptote (voir le paragraphe 3 du chap. 9) ; celui-ci est forcément d'altitude plus faible que celle de la surface de $N_e$ maximal.

En ne tenant compte que de l'effet des électrons libres, la traversée d'une couche ionosphérique donnerait, pour l'observateur $S$ sous celle-ci et dans le cadre d'un modèle de Terre plate avec une ionosphère stratifiée, une déviation totale $\chi_S$ nulle ; cette approximation serait acceptable quand la distance zénithale apparente $Z'$ n'est pas trop grande, donc on aurait une $\chi_S$ (*i.e.* la réfraction astronomique vue par $S$) quasi nulle en vertu de la formule de Simpson – relation (18) de [1] – où la réfractivité en $S$ est $\eta_S \cong 0$ vers 20 MHz. Ce n'est plus vrai dans un modèle à symétrie sphérique, la courbure des couches ionosphériques produisant pour $Z'$ assez grand des valeurs de $\chi_S$ très notables vers 20 MHz ; la formule de Laplace – relation (18) de [30] – n'est plus applicable, car l'altitude de l'ionosphère n'est plus assez petite devant le rayon terrestre, et surtout car les variations de $n$ avec l'altitude ne sont plus vraiment petites devant 1.

De plus, des effets de réfraction dus à de forts gradients horizontaux de $N_e$ sont bien connus des radioastronomes depuis les années 1950.

**5. Conclusion**

Les formules permettant de calculer $n$ en fonction de $\lambda_0$, $T$, $P$, $h$ et $x_c$ étant assez intriquées, son informatisation est appréciable. Plusieurs propositions existent, pour le domaine optique (de l'ultraviolet proche à l'infrarouge proche); nous recommandons celle de l'organisme américain *National Institute of Standards and Technology*, basée sur les formules de Ciddor, qui font le plus autorité actuellement :

[NIST Refractive Index of Air Calculator (based on Ciddor equation)](https://emtoolbox.nist.gov/Wavelength/Ciddor.asp)
https://emtoolbox.nist.gov/Wavelength/Ciddor.asp

Le même organisme propose aussi un calculateur basé sur une formule antérieure, ce qui permet d'analyser des travaux anciens :
NIST Refractive Index of Air Calculator (based on modified Edlén equation)
https://emtoolbox.nist.gov/Wavelength/Edlen.asp

## Références

[1] L. Dettwiller, « Panorama historique de l'étude de la réfraction astronomique : une histoire méconnue entre Optique, Mathématiques et Géodésie », *C. R. Phys.* **23** (2022), p. 13-62.
[2] B. Edlén, « The dispersion of standard air », *J. Opt. Soc. Am.* **43** (1953), p. 339-344.
[3] B. Edlén, « The refractive index of air », *Metrologia* **2** (1966), p. 71-80.
[4] G. Bönsch, E. Potulski, « Measurement of the refractive index of air and comparison with modified Edlén's formulae », *Metrologia* **35** (1998), p. 133-139.
[5] P. E. Ciddor, « Refractive index of air: new equations for the visible and near infrared », *Appl. Opt.* **41** (1996), p. 1566-1573.
[6] P. E. Ciddor, R. J. Hill, « Refractive index of air: 2. Group index », *Appl. Opt.* **38** (1999), p. 1663-1667.
[7] P. E. Ciddor, « Refractive index of air: 3. The roles of $CO_2$, $H_2O$, and refractivity virials », *Appl. Opt.* **41** (2002), p. 2292-2298.
[8] K. P. Birch, M. J. Downs, « An updated Edlen equation for the refractive index of air », *Metrologia* **30** (1993), p. 155-162.
[9] H. A. Lorentz, « Concerning the relation between the velocity of propagation of light and the density and composition of media », *Verh. Kon. Akad. Wetensch. Amsterdam* **18** (1878), p. 1-119, in *Collected papers*, t. 2, Nijhoff, La Hague, 1936.
[10] M. Bertin, J.-P. Faroux, J. Renault, *Électromagnétisme 4 – milieux diélectriques et milieux aimantés*, Dunod Université, Paris, 1984, voir p. 63-68.
[11] K. P. Birch, « Precise determination of refractometric parameters for atmospheric gases », *J. Opt. Soc. Am. A* **8** (1991), p. 647-651.
[12] H. Barrell, J. E. Sears, « The refraction and dispersion or air for the visible spectrum », *Phil. Trans. Roy. Soc. Lond. A* **238** (1939), p. 1-64.
[13] B. Edlén, « The refractive index of air », *Metrologia* **2** (1966), p. 71-80.
[14] J. C. Owens, « Optical refractive index of air: dependence on pressure, temperature and composition », *Appl. Opt.* **6** (1967), p. 51-59.
[15] K. E. Erickson, « Investigation of the invariance of atmospheric dispersion with a long-path refractometer », *J. Opt. Soc. Am.* **52** (1961), p. 777-780.
[16] F. E. Jones, « The refractivity of air », *J. Res. NBS* **86** (1981), p. 27-32.
[17] E. R. Peck, K. Reeder, « Dispersion of air », *J. Opt. Soc. Am.* **62** (1972), p. 958-962.
[18] J. G. Old, K. L. Gentili, E. R. Peck, « Dispersion of carbon dioxide », *J. Opt. Soc. Am.* **61** (1971), p. 89-90.
[19] P. Provost, J.-P. Provost, *Thermodynamique physique et chimique – Notions de flux et d'irréversibilité*, CEDIC / Nathan, Paris, 1984.
[20] R. C. Stone, « An accurate method for computing atmospheric refraction air », *PASP* **108** (1996), p. 1051-1058.
[21] R. S. Davis, « Equation for the Determination of the Density of Moist Air (1981/91) », *Metrologia* **29** (1992), p. 67-70.
[22] F. Arago, « Mémoire sur les affinités des corps pour la lumière et particulièrement sur les forces réfringentes des différents gaz », in *Œuvres complètes* (J.-A. Barral, éd.), t. 11, vol. 2, Gide, Paris, 1859, p. 702-732.
[23] F. Arago, « Mémoire sur la méthode des interférences appliquée à la recherche des indices de réfraction », in *Œuvres complètes* (J.-A. Barral, éd.), t. 10, Gide, Paris, 1858, p. 312-334.
[24] J. Jamin, « Mémoire sur l'indice de réfraction de la vapeur d'eau », *Ann. Chim. Phys.*, 3e série, **52** (1858), p. 171-188.
[25] J.-B. J. Delambre, *Astronomie Théorique et Pratique,* Tome Premier, Courcier, Paris, 1814.
[26] J.-B. Biot, F. Arago, « Mémoire sur les affinités des corps pour la lumière, et particulièrement sur les forces réfringentes des différens gaz », *Mém. Inst.* **7** (1806), p. 39-66.
[27] M. M. Colavita, M. R. Swain, R. L. Akeson, C. D. Koresko, R. J. Hill, « Effects of atmospheric water vapor on infrared interferometry », *Pub. Astron. Soc. Pacific* **116** (2004), p. 876-885.
[28] R. J. Mathar, « Calculated refractivity of water vapor and moist air in the atmospheric window at 10 μm », *Appl. Opt.* **43** (2004), p. 928-932.
[29] R. J. Mathar, « Refractive index of humid air in the infrared: Model fits », *J. Optics A: Pure Appl. Opt.* **9** (2007), p. 470-476.
[30] L. Dettwiller, « Propriétés remarquables de la réfraction astronomique dans une atmosphère à symétrie sphérique », *C. R. Phys.* **23** (2022), p. 63-102.
[31] L. Essen, K. D. Froome, « The Refractive Indices and Dielectric Constants of Air and its Principal Constituents at 24,000 Mc/s », *Proc. Phys. Soc. B* **64** (1951), p. 862-875.
[32] Symposium - International Astronomical Union, Refractional Influences in Astrometry And Geodesy, *Proceedings of the International Astronomical Union* **89** (1979)